\documentclass[aps,prc,twocolumn,longbibliography,superscriptaddress,amsmath,nofootinbib]{revtex4-1}

\usepackage[colorlinks, urlcolor=blue,linkcolor=blue, anchorcolor=blue, citecolor=blue]{hyperref}
\usepackage[utf8]{inputenc}
\usepackage{lineno}
\usepackage{amssymb}
\usepackage{amsmath}
\usepackage{amsfonts}
\usepackage{bm}
\usepackage{mathcomp}
\usepackage{graphicx}
\usepackage{subfigure}
\usepackage{multirow}
\usepackage[dvipsnames]{xcolor}
\usepackage[normalem]{ulem}

\hypersetup{bookmarksopen=true}
\graphicspath{{fig/}}

\newcommand{\bes}{{\sc BEShydro}}
\newcommand{\isd}{{\sc iS3D}}

\newcommand{\n}{n}
\newcommand{\ed}{e}

\newcommand{\peq}{p}

\newcommand{\ehat}{\hat e}
\newcommand{\qhat}{\hat q}

\newcommand{\eq}{{\,=\,}}

\begin{document}


\title{Response of a baryon-charged medium to energetic partons}

\author{Lipei Du}
\affiliation{Department of Physics, McGill University, Montreal, Quebec H3A 2T8, Canada}
\affiliation{Department of Physics, The Ohio State University, Columbus, Ohio 43210, USA}
\author{Ulrich Heinz}
\affiliation{Department of Physics, The Ohio State University, Columbus, Ohio 43210, USA}

\begin{abstract}
We explore the response to energetic partons of a baryon-charged medium produced in low energy heavy-ion collisions in which the partonic energy loss rate is expected to depend on both temperature and baryon chemical potential. The energy and momentum deposited by the partons are described by dynamical sources added to hydrodynamic equations of motion. We study the distortions of various hydrodynamic quantities, especially the energy and baryon densities, induced by energetic partons plowing through the medium. By studying the contribution from this medium response to the emission spectra of several identified hadron species we identify qualitative differences between the jet-induced modifications for mesons and net protons. 
\end{abstract}
\date{\today}
\maketitle

\section{Introduction}

Mapping the QCD phase diagram, where a hypothetical QCD critical point serves as a landmark, is one of the primary goals of the high energy nuclear community \cite{Bzdak:2019pkr,An:2021wof}. At high enough temperatures, confined quarks and gluons can be liberated from hadrons and quark-gluon plasma (QGP) is formed, a state of matter which existed during the first few microseconds of the early universe. This state can be recreated in high energy heavy-ion collisions, and several signatures, including collective flow, strangeness enhancement and jet quenching, have contributed to the discovery of the QGP \cite{Gyulassy:2004zy}. This exotic matter has been found to be the least viscous liquid, with the highest temperature, the smallest size and the highest rotational speed (${\sim\,}10^{22}\,$s$^{-1}$) ever created on earth \cite{Heinz:2013th,Schenke:2021mxx,Becattini:2020ngo}. 

The QCD critical point corresponds to a second-order phase transition between the phases of QGP and hadron gas. Confirming its existence and constraining its location in the phase remain open problems of topical importance \cite{Bzdak:2019pkr,An:2021wof}. Furthermore, understanding the onset of deconfinement (by varying the collision energy) and the onset of the formation of a fireball with collective dynamical behavior (by varying the size of the colliding nuclei) are additional topics of great interest. Systematic studies of collision energy and system size dependencies may help us to figure out how a medium that manifests itself as a strongly coupled QGP liquid at a coarse spatial resolution of order ${\sim\,}1/T$ emerges from the weakly coupled fundamental quark and gluon degrees of freedom that characterize the strong interaction at much shorter length scales \cite{Busza:2018rrf}. A key question such searches are to answer is the following: What is the smallest amount of strongly interacting matter that can still be described hydrodynamically as a droplet of QGP liquid?

Jets interacting with QGP matter created in heavy-ion collisions are characterized by at least two distinct length scales: a short one ${\sim\,}1/Q$ related to the virtuality $Q^2$ of the hard parton(s) when first created in the collision, and a longer one ${\sim\,}1/T$, determined by the temperature of the QGP medium and characterising the jet quenching parameter $\hat{q}$ which describes transverse momentum transfer between the hard parton and the thermal medium as the jet ploughs through the QGP \cite{Qin:2015srf, Cao:2020wlm}. This makes them good candidates for exploring the emergence of nonperturbative thermal and collective behavior at low momentum scales from perturbative interactions at high momentum scales. This is the motivation behind studying the hydrodynamic response to jet quenching in this work. While many such studies already exist for high collision energies where the QGP created in the collision is essentially baryon neutral, we here explore the jet-medium response in lower-energy collisions where the medium has large net baryon density and where the thermalization of jets generated by hard partons, especially hard quarks and antiquarks, can lead to interesting baryon diffusion effects. 

The jet production cross section has a strong collision energy dependence and therefore high-virtuality jets become extremely rare in low-energy heavy-ion collisions such as those studied in the Beam Energy Scan (BES) program at the Relativistic Heavy Ion Collider (RHIC) \cite{Bzdak:2019pkr}. Nevertheless, energetic primordial partons can still be produced, albeit at low rates, and when they traverse the QGP medium and interact with its constituents they exchange energy and momentum with the medium, leading to enhanced soft particle production in the final state along its direction of propagation \cite{Horvat:2013lza, Adamczyk:2017nof}. Although such jets can not be reliably reconstructed from the final-state hadrons because their decay products are embedded in a large soft-particle background, they should still manifest as an excess of hadrons with larger than average momenta ($p_T{\,\sim\,}$several GeV) above a background of soft particles with $p_T\lesssim\,1$\,GeV. Jet quenching will move that excess to smaller momentum, leading to a localized enhancement of the soft background itself, and this effect should be stronger in central than in peripheral collisions, due to the longer distances travelled by the energetic partons in the larger fireballs created in the former.  

A systematic measurement of jet-quenching effects at BES energies has been recently reported by the STAR Collaboration \cite{Adamczyk:2017nof}, and transport model calculations were carried out but so far failed to reproduce the measurements \cite{Horvat:2013lza, Horvat:2017}. The exploratory work presented here is a limited attempt, using hydrodynamics, to provide a framework for qualitative theoretical predictions to aid the analysis and interpretation of these data. Our focus is on the response of a {\it baryon-charged} medium to energetic partons.\footnote{%
    Following custom we will occasionally use the term ``jet'' to refer to such energetic partons, without implying that their fragments manifest in a clearly recognizable spray of hard final-state particles.}
The main reason for our work remaining exploratory is the continued lack of a fully calibrated 3-dimensional hydrodynamic model of the evolution of the baryon-charged fireballs created at BES energies. While including the effects of finite net baryon density in the medium, our analysis of the medium response to energetic partons will therefore based on a highly simplified background medium profile. These simplifications include the assumption of an ideal (i.e.\ non-dissipative) background fluid. The baryonic response described here will therefore be restricted to jet-induced ideal net baryon flow and neglect effects caused by baryon diffusion.   

Noting that at low beam energies the interaction between energetic partons and medium constituents may not be well described by a weakly coupled theory we will employ  calculations for strongly coupled systems based on gauge/gravity duality for key transport coefficients. While such holographic models do not admit a quasi-particle interpretation in terms of particle-like jet constituents, they do show an enhancement of the jet quenching parameter $\qhat$ near the pseudo-critical temperature $T_{pc}$ for the quark-hadron phase transition which further increases with increasing baryon chemical potential \cite{Rougemont:2015wca}. This feature is qualitatively shared by weakly coupled theories that do admit a quasi-particle picture \cite{Liao:2008dk} in which it can be understood as a consequence of the relation $\qhat/T^3\propto1/(\eta/s)$ \cite{Majumder:2007zh, Muller:2021wri} between the jet quenching parameter $\qhat$ and the specific shear viscosity $\eta/s$ which has a minimum near $T_{pc}$ \cite{Csernai:2006zz}. Such an enhancement in jet quenching near the phase transition is of great interest since it opens the possibility to gain additional insight into the confinement/deconfinement process by studying the interaction between energetic partons and the QGP medium produced in low beam energy collisions.

In this work we explore the distortions in the baryon sector of a baryon-charged medium as it responds to single and correlated pairs of hard partons. We use an energy loss rate that depends on both temperature and chemical potential, constructed from holographic calculations \cite{Rougemont:2015wca}. The medium excitations caused by a traversing parton(s) are assumed to be instantaneously thermalized and to evolve hydrodynamically together with the ideal fluid background medium, using hydrodynamic equations with dynamical sources. With this setup we study the jet-induced corrections to the final-state distributions of identified particles, in particular protons and anti-protons which are sensitive to the flow and diffusion of baryon charge currents.

This work is organized as follows: Sec.~\ref{sec:model} describes our model setup, including the hydrodynamic evolution of the baryon-charged medium, the dynamical sources describing the energy-momentum deposition from partons, and the calculation of the resulting distortion in the final particle distributions. In Sec.~\ref{sec:results} we show how the medium gets distorted by a single parton and by a correlated parton pair. We explore differences between the final-state distributions of baryons with different baryon number  arising from differences in the corrections caused by the quenching of energetic partons. A summary and conclusions are offered in Sec.~\ref{sec:conclusion}. 

\section{Model and setup}
\label{sec:model}

In this section we describe our model setup, starting with the hydrodynamic evolution of the background medium at non-zero net baryon density in Sec.~\ref{sec:hydro}, followed by the modelling of the medium modification via source terms describing the deposition of energy, momentum and baryon number by penetrating partons in Sec.~\ref{sec:source}. The energy loss rate of such partons, needed for the constructing the dynamical sources, as obtained from holographic models is discussed in Sec.~\ref{sec:energyloss}. Formulae needed for calculating the corrections induced by energetic partons to the final particle distributions and the fluid's vorticity are compiled in Sec.~\ref{sec:particle} and Sec.~\ref{sec:vortex}.

\subsection{Evolution of the background medium}
\label{sec:hydro}

Most studies so far on jet-medium interaction have been carried out for ultra-relativistic heavy-ion collisions in which the collision fireball is baryon-charge neutral and conserved charge currents can be ignored. In this study, to explore baryon-charged medium response to energetic partons, we complement energy-momentum conservation by also including the conservation law of net baryon charge, while continuing to ignore strangeness and electric charge currents. 

In the absence of parton energy loss, the conservation laws for energy-momentum and the baryon charge are
\begin{equation}
    d_{\mu}T^{\mu\nu} = 0\;,\quad d_{\mu}N^{\mu} = 0\,, \label{eq:conslaws}
\end{equation}
where $d_\mu$ denotes the covariant derivative in a generic coordinate system. Here $T^{\mu\nu}$ and $N^{\mu}$ are the energy-momentum tensor and net baryon current, respectively, with the following hydrodynamic decompositions:
\begin{equation}
    T^{\mu\nu} = \ed u^{\mu}u^{\nu}- (\peq+\Pi) \Delta^{\mu\nu}+\pi^{\mu\nu}\;,\quad N^{\mu} = \n u^{\mu}+n^\mu\,.\label{eq:decomp}
\end{equation}
$u^{\mu}$ is the flow velocity, $\ed$ the energy density in the local rest frame (LRF), $\peq$ the thermal pressure, and $\n$ the LRF baryon density. $u^\mu u^\nu$ and $\Delta^{\mu\nu} \equiv g^{\mu\nu} - u^{\mu}u^{\nu}$ are projectors on the temporal and spatial directions in the LRF. We use the Landau definition of the LRF where the flow velocity $u^\mu$ (normalized by $u^\mu u_\mu=1$) is defined as the timelike eigenvector of $T^{\mu\nu}$, with eigenvalue~$\ed$: $T^{\mu\nu}u_\nu = \ed u^\mu$. The shear stress tensor $\pi^{\mu\nu}$, bulk viscous pressure $\Pi$ and baryon diffusion current $n^\mu$ are dissipative flows describing off-equilibrium effects.

While these dissipative flows are phenomenologically important for quantitative studies, Refs.~\cite{Yan:2017rku,Yan:2017gmv} found that they can be neglected in the calculation of excess soft particle production from quenching jets. Since we are predominantly interested in the medium modification effects caused by the jets and not aiming for a quantitative calibration of the unmodified background medium dynamics, we will use the ideal fluid approximation for the background evolution:  
\begin{eqnarray}
    Dn &=& -n\theta\;,\label{eq-n}\\
    De &=& -(e+p)\theta\;,\label{eq-e}\\
    Du^\mu &=& \nabla^\mu p/(e+p) \;.\label{eq-u}
\end{eqnarray}
Here $D{\,\equiv\,}u_\mu d^\mu$ stands for the time derivative in the LRF, $\theta{\,\equiv\,}d_\mu u^\mu$ is the scalar expansion rate, and $\nabla^\mu{\,\equiv\,}\partial^{\langle\mu\rangle}$ (where generally $A^{\langle\mu\rangle}{\,\equiv\,}\Delta^{\mu\nu}A_\nu$) denotes the spatial gradient in the LRF. These equations have simple intuitive interpretations; for example, Eq.~\eqref{eq-u} shows that fluid acceleration is driven by the pressure gradients working against the inertia provided by the enthalpy density $w\equiv\ed{\,+\,}p$. The scalar expansion rate $\theta$ controls the dilution of the LRF charge density $n$ (Eq.~(\ref{eq-n})) and the energy density $\ed$ (Eq.~(\ref{eq-e})), with an additional loss of local energy density caused by work done by the pressure $p$ which converts thermal energy density $e$ into kinetic energy associated with the collective flow.

Following the previous studies \cite{Yan:2017rku, Yan:2017gmv} on jet-medium interaction, we here employ the semi-analytic ideal Gubser flow profile \cite{Gubser:2010ze, Gubser:2010ui} for the evolution of the baryon-charged background medium as described in Refs.~\cite{Du:2020bxp, Du:2019obx} (see also Chapter~5 of Ref.~\cite{Du:2021fyr}) to which we refer the reader for additional details. Gubser flow describes longitudinally boost-invariant and azimuthally symmetric transverse expansion of a conformal system that appears static when expressed in the curvilinear Gubser coordinates parametrizing de Sitter space \cite{Gubser:2010ze, Gubser:2010ui}. When expressed in Milne coordinates $(\tau,r,\phi,\eta_s)$ the flow profile is independent of space-time rapidity $\eta_s$ and the azimuthal angle $\phi$, reading
\begin{align}
u^\tau(\tau, r) &= \cosh\kappa(\tau,r)\;,\quad u^\phi(\tau, r) = u^\eta(\tau, r) = 0\;,
\label{eq-gubser-ut}
\\
u^{x}(\tau, r) &= \frac{x}{r}\sinh\kappa(\tau,r)\;,\ 
u^{y}(\tau, r) = \frac{y}{r}\sinh\kappa(\tau,r)\;,
\label{eq-gubser-uxy}
\end{align}
where $\kappa(\tau,r)$ is the flow rapidity, corresponding to the longitudinally boost-invariant, azimuthally symmetric transverse flow velocity 
\begin{equation}
    v(\tau,r)=\tanh \kappa(\tau,r) \equiv 
    \frac{2q^2\tau r}{1+q^2\tau^2+q^2r^2}\;.
\label{eq-gubser-kappa}
\end{equation}
Gubser flow requires a conformal Equation of State (EoS) for which we use an ideal gas of massless quarks and gluons,
\begin{equation}
    e = 3p = f_*(\alpha) T^4\,,\quad n = \alpha g_*(\alpha) T^3\,,\label{eq:eos}
\end{equation}
where $f_*(\alpha)$ and $g_*(\alpha)$ are unitless coefficients depending on the normalized baryon chemical potential $\alpha\equiv\mu/T$ \cite{Du:2020bxp}.\footnote{%
    This is EOS3 in Ref.~\cite{Du:2019obx}.}
The corresponding temperature profile  in Milne coordinates is given by
\begin{equation}
    T(\tau,r) = \frac{C}{\tau}\frac{(2q\tau)^{2/3}}{\bigl[1+2q^2(\tau^2{+}r^2)+q^4(\tau^2{-}r^2)^2\bigr]^{1/3}}\,,
\label{eq-gubser_temp}
\end{equation}
where $q$ is an arbitrary energy scale that controls the transverse size of the system, and $C$ is a constant of integration \cite{Gubser:2010ze} (cf.~Fig.~\ref{fig:jet_illustration}b below). With the EoS \eqref{eq:eos} the profiles of the other thermodynamic quantities are easily obtained \cite{Du:2019obx}.

By using Gubser expansion for the background medium, we trade phenomenological precision for analytical control over the background, without sacrificing important qualitative features of fireballs created in heavy-ion collisions, such as simultaneous longitudinal and transverse expansion. Although Gubser flow requires conformal symmetry of the background medium we shall not insist the conformality of other aspects in the setup related to the interaction of jets with that medium. In practice, we shall simulate the medium evolution using the \bes{} code \cite{Du:2019obx} with initial conditions given by the Gubser profiles. The distortions induced by the partons can be obtained easily by subtracting the analytical Gubser background from the perturbed numerical results --- this is one of the benefits of using the Gubser flow.

\subsection{Hydrodynamics with dynamical sources}\label{sec:source}

An energetic parton traveling through the background medium
distorts the latter by interacting with it. Different approaches for describing the medium response are reviewed in Refs.~\cite{Tachibana:2019hrn, Luo:2021iay, Qin:2015srf, Cao:2020wlm}. In the weak-coupling approach energy and momentum are exchanged through partonic scatterings between jet partons and thermal particles in the medium. In a strongly-coupled picture partons with energies below some threshold are assumed to be fully absorbed by the medium, becoming part of the hydrodynamically evolving fluid. Other studies combine the two approaches (see, e.g., Refs.~\cite{Chen:2017zte, Chen:2020tbl, Yang:2021qtl, Yang:2022nei, Tachibana:2020mtb, Tachibana:2020atn}).

In this work, we assume that the energy-momentum and baryon charge deposited by the energetic parton are  instantaneously thermalized such that their subsequent evolution can be described hydrodynamically. We account for them by adding dynamical source terms to the hydrodynamic equations \eqref{eq:conslaws}:
\begin{equation}
    d_\mu T^{\mu\nu}(x) = J^\nu(x)\,,\quad
    d_\mu N^{\mu}(x) =\rho_B(x)\,,
\label{eq:hydro_source}
\end{equation}
with
\begin{equation}
    J^\nu(x)=-d_\mu T^{\mu\nu}_\mathrm{p}(x)\,,\quad
    \rho_B(x) = -d_\mu N^{\mu}_\mathrm{p}(x)\,.
\label{eq:sources}
\end{equation}
Here $T^{\mu\nu}_\mathrm{p}(x)$ and $N^{\mu}_\mathrm{p}(x)$ denote the energy-momentum tensor and net baryon current of the energetic parton(s) traversing the medium. Eqs.~(\ref{eq:hydro_source},\ref{eq:sources}) ensure the local conservation of energy-momentum and net baryon number for the combined system consisting of the medium and the parton(s): $d_\mu (T^{\mu\nu}+T^{\mu\nu}_\mathrm{p})=0$ and $d_\mu (N^{\mu}+N^{\mu}_\mathrm{p})=0$. When the parton traverses the medium and induces excitations (described by $J^\nu$ and $\rho_B$), their associated energy, momentum and baryon charge become part of the hydrodynamically evolving medium described by $T^{\mu\nu}$ and $N^{\mu}$. We ignore the quantization of baryon number and treat the baryon source term $\rho_B(x)$ as a continuous function.

To solve Eqs.~\eqref{eq:hydro_source} numerically, we use \bes{} with a dynamical source module  \cite{Du:2018mpf,du2020ds}, developed for incorporating dynamical initialization which is necessary for describing the initial stage of collisions at low beam energies. Some other studies (e.g. Refs.~\cite{Yan:2017gmv,Yan:2017rku,Casalderrey-Solana:2020rsj}) assume that the dynamical sources are small perturbations compared to the medium, using linear response theory to solve Eqs.~\eqref{eq:hydro_source}:
\begin{eqnarray}
    D\delta n+d_\mu(n\delta u^\mu)&=&\rho_B\,,\label{eq:hydro_source_pert_n}\\
    D\delta e+\delta w\theta +d_\mu(w\delta u^\mu)+w\delta u^\mu Du_\mu&=&-u^\mu J_\mu\,,\label{eq:hydro_source_pert_e}\\
    \delta wDu^\nu+(Dw+w\theta)\delta u^\nu+&&\nonumber\\
    w\delta u^\mu d_\mu u^\nu+\nabla^\nu \delta p+wD\delta u^\nu &=& J^\nu\,.
\label{eq:hydro_source_pert_u}
\end{eqnarray}
These linearized equations can provide intuitive guidance for understanding the full numerical results presented in Sec.~\ref{sec:singlejet} below. For example, Eq.~\eqref{eq:hydro_source_pert_n} shows that both the baryon source term $\rho_B$ and the flow distortion $\delta u^\mu$ induced by the fast parton contribute to the perturbation $\delta n$ of the baryon density.

To construct the dynamical sources in Eqs.~(\ref{eq:hydro_source},\ref{eq:sources}) for a jet shower one can employ a kinetic description for $T^{\mu\nu}_\mathrm{p}$ and $N^{\mu}_\mathrm{p}$ in terms of the phase-space distributions of the jet partons \cite{Tachibana:2017syd}. For the energy-momentum source current $J^\mu$ we use an expression given in Refs.~\cite{Yan:2017gmv,Yan:2017rku}:
\begin{equation}
    J^\mu(t, \boldsymbol x) = \frac{dE}{dt}\, u^\mu_\mathrm{p}\, n_\mathrm{p}(t, \boldsymbol x; \boldsymbol x_\mathrm{p})\,.
\label{eq:jterm}
\end{equation}
Here $E$ is the energy of the jet parton, $u^\mu_\mathrm{p}=(1,\boldsymbol v_\mathrm{p})$ is its light-like four-velocity, $n_\mathrm{p}(t, \boldsymbol x; \boldsymbol x_\mathrm{p})$ the number density distribution of the partons, and $dE/dt$ the energy loss rate (see Sec.~\ref{sec:energyloss} below).\footnote{%
    As pointed out in Refs.~\cite{Yan:2017gmv,Yan:2017rku}, Eq.~(\ref{eq:jterm}) ignores the induced jet-medium interaction arising from the transverse momentum broadening in the evolution of the jet source which is suppressed by a factor of order $O(T/p_\mathrm{jet})$ when $p_\mathrm{jet}\gg T$.}
$\boldsymbol x_\mathrm{p}(t) = \boldsymbol x_0+(t-t_0) \boldsymbol v_\mathrm{p}$ is the trajectory of the light-like parton, with $(t_0, \boldsymbol x_0$) denoting the Cartesian coordinates of its initial point of creation.
We here focus on jet partons emitted in the plane transverse to the beam at midrapidity. Correspondingly, at $z=0$, $\boldsymbol v_\mathrm{p} = (\boldsymbol v_{\mathrm{p}\perp},v_z) = (\boldsymbol v_{\mathrm{p}\perp},0)$. To implement the current (\ref{eq:jterm}) in \bes{} we must express its components in
Milne coordinates $(\tau,x,y,\eta_s)$. Noting that $t=\tau$ at $z=0$, we substitute $dE/dt\mapsto dE/d\tau$. Incorporating this modified current in a longitudinally boost-invariant background medium without the restriction to $z=0$ extends the boost-invariance also to the source term, effectively replacing the ``pin"-like jet parton by a ``knife" oriented along the $z$ axis that cuts the medium in transverse direction \cite{Chaudhuri:2005vc,Yan:2017rku}. This problem is easier to study than the medium response to a localized parton that is created at mid-rapidity with non-zero longitudinal momentum which breaks the longitudinal boost-invariance \cite{Casalderrey-Solana:2020rsj}.

For $n_\mathrm{p}(t, \boldsymbol x; \boldsymbol x_\mathrm{p})$ we use the following boost-invariant Gaussian smearing kernel:
\begin{equation}
\label{eq:smearing}
    n_\mathrm{p}(\tau, \boldsymbol x_\perp; \boldsymbol x_\mathrm{p\perp}) = \frac{1}{\tau}\frac{1}{2\pi\sigma^2} \exp\left[-\frac{(\boldsymbol x_\perp - \boldsymbol x_\mathrm{p\perp})^2}{2\sigma^2}\right]\,,
\end{equation}
where $\boldsymbol x_\mathrm{p\perp}\equiv(x_\mathrm{p},y_\mathrm{p})$ is the parton's transverse position perpendicular to the beam direction. Eq.~(\ref{eq:smearing}) is normalized as $\int \tau d\tau d^2\boldsymbol x_\perp\, n_\mathrm{p}(\tau, \boldsymbol x_\perp; \boldsymbol x_\mathrm{p\perp})=1$ and independent of space-time rapidity. 

In this work we shall assume that no baryon charge is deposited by the energetic parton:
\begin{equation}\label{eq:bsource}
    \rho_B(t, \boldsymbol x) = 0\,.
\end{equation}
This is a crude way of dealing with the issue that even a quark parton does not deposit its baryon charge of $1/3$ continuously into the medium. Of course, energetic partons (quarks or gluons) can radiate gluons which then split into $q$-$\bar q$ pairs, producing additional quark partons that can annihilate on anti-quarks from the medium, thereby changing the local baryon density of the medium. Our assumption $\rho_B=0$ implies that, even if the jet parton is dressed by a cloud of $q$-$\bar q$ pairs, this cloud does not dissolve into the medium. Ref.~\cite{Chesler:2008uy} pointed out that in strongly coupled ${\cal N}=4$ supersymmetric Yang-Mills theory the baryon density of an energetic parton can remain highly localized over long distances. Additional work on baryon doping of the medium by energetic partons is required to explore to what extent this finding carries over to real QCD.
 
\subsection{Energy loss rate}
\label{sec:energyloss}

Within a perturbative QCD picture both elastic and inelastic collisions between the parton and the medium constituents can induce loss of parton energy. These processes are conventionally encoded in the transport coefficients $\ehat\equiv dE/dt$, which describes the energy loss rate due to elastic collisions with medium constituents, and $\qhat\equiv d(\Delta p_\perp)^2/dt$ which parametrizes the transverse momentum broadening of the jet shower and encodes radiative energy loss via gluon emission induced by inelastic scattering in the medium \cite{Baier:1996sk}. In a medium which is close to local equilibrium the fluctuation-dissipation theorem \cite{Moore:2004tg} relates these two coefficients by
\begin{equation}\label{eq:ehatqhat}
    \ehat \propto \qhat/T\,,
\end{equation}
with prefactor to be discussed below.

\begin{figure}[!tbp]
    \centering
    \hspace*{-0.30cm}\includegraphics[width=1.06\linewidth]{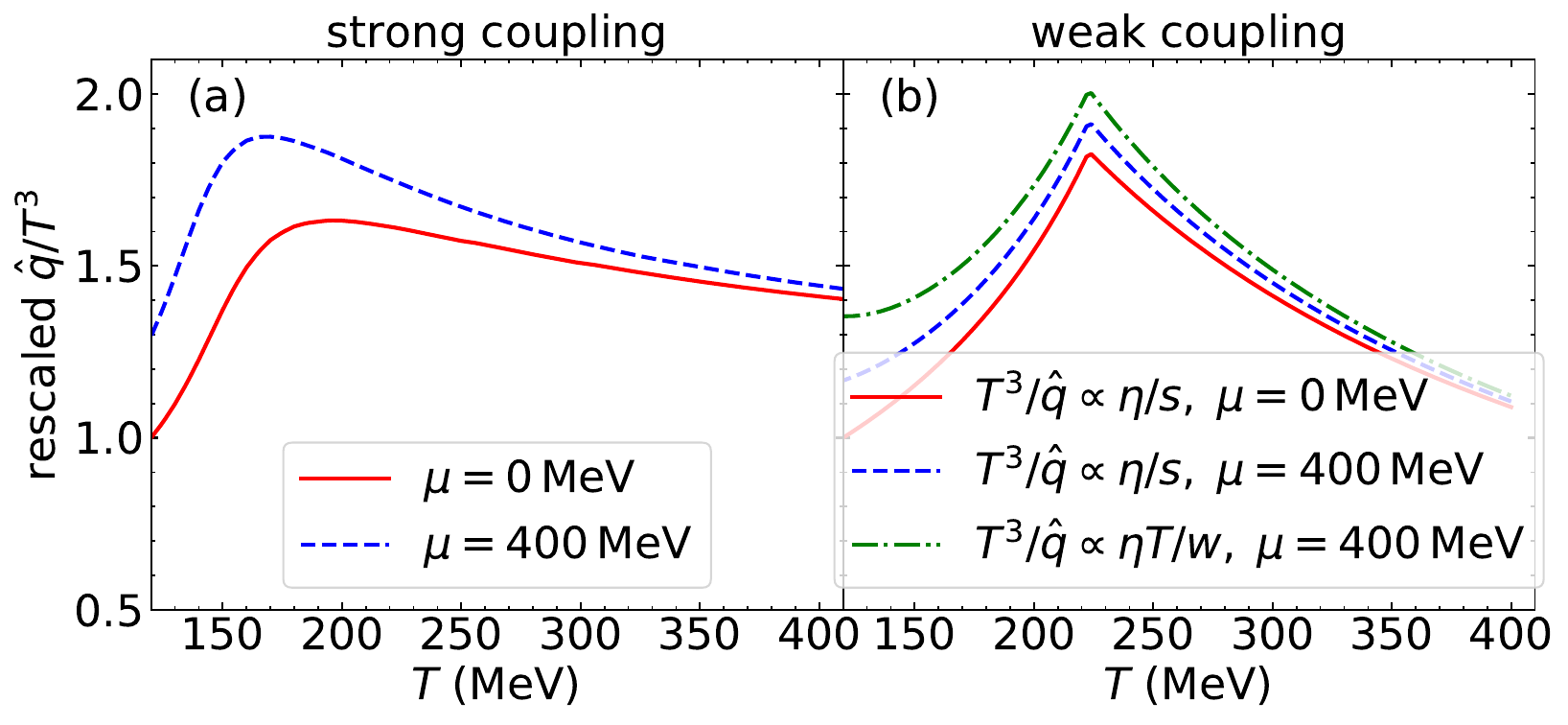}
    \caption{Temperature dependence of $\qhat/T^3$, plotted as a multiple of its value at $(T,\mu)=(120,0)$\,MeV, for various values of the chemical potential $\mu$, within the (a) strongly \cite{Rougemont:2015wca} and (b) weakly coupled \cite{Majumder:2007zh,Muller:2021wri} pictures. In (b) the red solid and blue dashed lines use the relation $T^3/\qhat\propto\eta/s$ derived at $\mu=0$ in \cite{Majumder:2007zh} while the green dash-dotted line generalizes this expression to nonzero $\mu$ by  substituting $sT=w$ (which holds at $\mu=0$) and using  $T^3/\qhat\propto\eta{}T/w$ at nonzero chemical potential. The specific shear viscosity $(\eta/s)(T)$ at $\mu\eq0$ was taken from Refs.~\cite{Everett:2020xug,Everett:2020yty} --- see text for details.
    \vspace*{-3mm}
    }
    \label{fig:qhat}
\end{figure}

In this work we use Eq.~\eqref{eq:ehatqhat} to obtain the chemical potential dependent $\ehat(\mu, T)$ from a calculation of $\qhat(\mu, T)$ within a holographic model \cite{Rougemont:2015wca}. The authors of Ref.~\cite{Rougemont:2015wca} first tuned model parameters to fit Lattice QCD calculations for the EoS and the baryon susceptibility near the crossover phase transition at $\mu\eq0$, and then calculated the pressure and the speed of sound at non-zero chemical potential for $\mu \leq 400\,$MeV where good quantitative agreement with corresponding Lattice QCD results was found \cite{Rougemont:2015wca}. The tuned model is then used to calculate $\qhat(T,\mu)$. In Fig.~\ref{fig:qhat}a we plot the ratio $\qhat(\mu, T)/\qhat_\mathrm{CFT}$ as a function of temperature for two values of the chemical potential, where the conformal denominator is given by \cite{Liu:2006ug,Liu:2006he}
\begin{equation}\label{eq:qcft}
    \qhat_\mathrm{CFT} = \frac{\pi^{3/2}\Gamma(\frac{3}{4})}{\Gamma(\frac{5}{4})}\sqrt{\lambda}\,T^3\,,
\end{equation}
with $\lambda$ being the 't Hooft coupling. Ref.~\cite{Rougemont:2015wca} estimates that $\qhat\approx27.22\, T^3$ for light quarks at $(T,\mu)\eq(398,0)$\,MeV, using $\lambda\eq7.8$. On the other hand, in a weakly coupled approach one obtains at $\mu=0$ \cite{Majumder:2007zh,Muller:2021wri} 
\begin{equation}\label{eq:qhatetas}
   \qhat\approx1.25\, T^3/(\eta/s)
\end{equation}
which, using $\eta/s\eq1/4\pi$ \cite{Policastro:2001yc, Kovtun:2003wp}, yields $\qhat \approx 16.76\, T^3$ (this expression is used in Ref.~\cite{Yan:2017gmv}). One sees that at $\mu=0$ the strongly and weakly coupled approaches yield values for $\qhat/T^3$ that agree within a factor 2.\footnote{%
    Accounting for the different number of degrees of freedom in QCD and the holograhic model actually helps to further improve the agreement \cite{Liu:2006he}.}

In Fig.~\ref{fig:qhat} the differences between the strongly and weakly coupled results at $\mu=0$ are divided out at the reference point $(T,\mu)=(120,0)\,$MeV, in order to focus on the differences in magnitude and shape of $\qhat/T^3$ between the two approaches as functions of both $T$ and $\mu$. Note that both panels cover identical ranges. In the weakly coupled approach shown in panel (b), the cusp at $T\approx 225$\,MeV results from the temperature dependence of the specific shear viscosity at $\mu=0$, $(\eta/s)(T,\mu{=}0)$, extracted from the Bayesian model calibration reported in \cite{Everett:2020xug, Everett:2020yty} using experimental data collected at the RHIC and LHC complexes, which features a minimum at this temperature. In the weakly coupled approach we must make a decision on how to extrapolate the factor $\eta/s$ into the region of non-zero chemical potential. In principle, both $\eta$ and $s$ depend on $\mu$. If we ignore the former, as suggested in Ref.~\cite{Denicol:2013nua} based on a calculation in the hadron resonance gas phase, and in the factor $\eta/s$ simply substitute $s(T,0)$ by $s(T,\mu)$ obtained from the conformal EoS in Eqs.~(\ref{eq:eos}), we obtain the blue-dashed line in Fig.~\ref{fig:qhat}b. Alternatively, we can take the identity $sT=e+p=w$ that holds at $\mu=0$ and substitute in the factor $\eta/s$ the entropy density $s$ by $w/T$ \cite{McLaughlin:2021dlq}.\footnote{%
    The use of $w=e+p=sT+\mu n$ in place of simply $sT$ is also suggested by Refs.~\cite{Liao:2009gb, Denicol:2013nua} where it was argued that $\eta/w$ is a better indicator than $\eta/s$ for the fluidity of a liquid at nonzero chemical potential.}
At non-zero $\mu$ where $w/T{\eq}s+\mu n/T$ this leads to a significant increase in $w$ and therefore of $\qhat/T^3 \propto w/T\eta$, as reflected in the green dash-dotted line in Fig.~\ref{fig:qhat}b. Finally, comparing panels (a) and (b), we note that, in spite of quantitative differences in detail, the strongly and weakly coupled approaches share many qualitative features, even though the former doesn't even admit the concept of quasiparticle constituents for the medium. The phenomenological relevance of the peak in the normalized energy loss near the phase transition was pointed out in Refs.~\cite{Liao:2008dk, Li:2014hja, Xu:2014tda, Xu:2015bbz, Renk:2014nwa}.

\subsection{Model parameters}
\label{sec:parameters}

To simulate the medium response to energetic partons, we shall use the following parameters. We set the initial temperature at the center of the fireball to $T\eq370\,$MeV \cite{Burke:2013yra}. Correspondingly, we use $C\approx3.24\,$fm$^{-1}$, together with $q^{-1}\eq4.3$\,fm,
for the Gubser profile in Eq.~\eqref{eq-gubser_temp} \cite{Gubser:2010ui, Gubser:2010ze}. We start the hydrodynamic evolution at the initial time $\tau_0\eq1\,$fm$/c$. To set the baryon density we use $\alpha\eq\mu/T\eq0.2$ for the background medium --- a value that corresponds to a collision energy $\sqrt{s_\mathrm{NN}}$ somewhat below 200\,GeV \cite{Du:2020bxp} and is large enough to give non-negligible baryonic effects but doesn't require us to extrapolate too far into the finite-$\mu$ direction. We use a freeze-out temperature $T_f\eq148\,$MeV motivated by a Bayesian inference \cite{Bernhard:2016tnd}.\footnote{%
    While the value extracted in \cite{Bernhard:2016tnd} should be interpreted as the chemical freeze-out temperature, we here use it also, for simplicity and because the evolution model used here lacks a hadronic cascade module to handle kinetic freeze-out, to define the kinetic freeze-out point.}
With the available analytical solution of Gubser flow, we estimate that the initial radius of the fireball region with a temperature above $T_f$ is $r\approx7.7$\,fm, and that the time needed for the fireball's center to drop to the freeze-out temperature is $\tau_f\approx4.3\,$fm$/c$.\footnote{%
    Note that this estimated lifetime is shorter than that extracted from more realistic simulations \cite{Everett:2020xug, Everett:2020yty} because Gubser flow features large initial transverse flow which helps to dilute the fireball much more rapidly.
}
We can then estimate how far the light-like parton can travel during this time period and use a large enough grid to contain it in the simulation. For the source profile \eqref{eq:smearing} we use a smearing width $\sigma\eq0.2\,$fm, and $\ehat\eq\qhat/(8T)$ for Eq.~\eqref{eq:ehatqhat}.\footnote{%
    We note that Refs.~\cite{Yan:2017gmv,Yan:2017rku} set $\ehat\eq\qhat/(4T)$. We are here using a smaller energy loss rate so that the source terms are small enough for linear response theory to be applicable. This is required for some of the discussion in Sec.~\ref{sec:dijet} below. Although using a larger energy loss rate would enhance the medium response quantitatively we do not expect it to cause qualitative changes.} 
We make the implicit assumption that the original energy of the jet parton is sufficiently large that this simple expression holds throughout its propagation through the medium.

\begin{figure}[!tbp]
    \centering
    \hspace{-0.5cm}\includegraphics[width=0.24\textwidth]{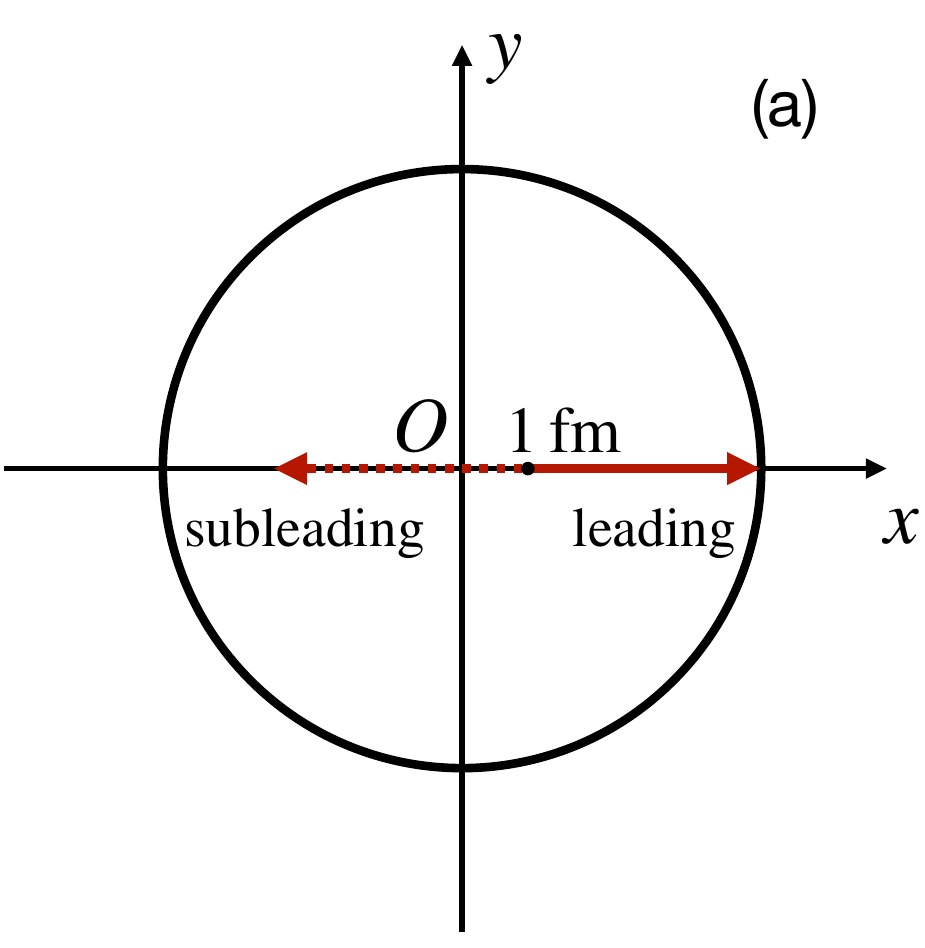}\hspace{0.1cm}\includegraphics[width=0.25\textwidth]{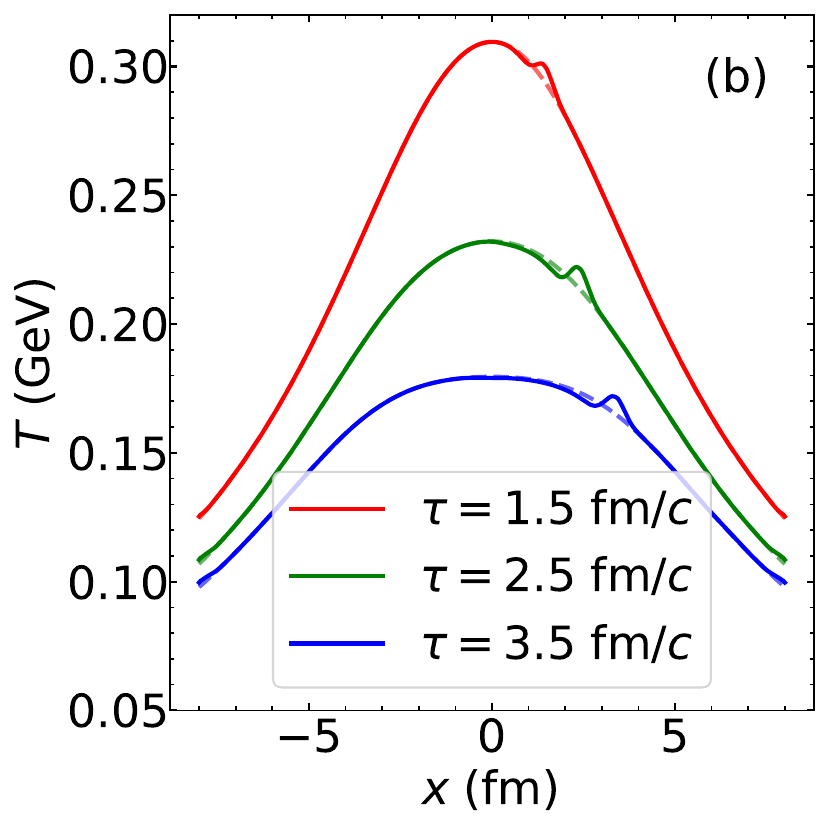}
    \caption{(a) Illustration of the fireball in the transverse plane at mid-rapidity, with two energetic partons moving in opposite directions, starting at $(x,\,y)=(1,\,0)$\,fm. The solid red arrow pointing right indicates the leading parton, whereas dashed red arrow pointing left is the corresponding subleading parton. (b) Temperature distributions along the $x$-axis at $y=0$ at different times. Broken lines are for the ideal Gubser flow background without any energetic partons, whereas solid lines include the contribution from a leading parton moving in $+x$ direction (solid arrow in (a)).
    \label{fig:jet_illustration}
    \vspace*{-3mm}
    }
\end{figure}

An illustration for the evolution of the temperature profile using the above setup is shown by the dashed curves in Fig.~\ref{fig:jet_illustration}b. The solid curves correspond to the distorted profile that includes the energy-momentum contributed to the medium by a single leading parton plowing through the fireball in $x$-direction, as represented by the solid red arrow pointing in $+x$ direction in Fig.~\ref{fig:jet_illustration}a. A more detailed discussion will follow below. 

\subsection{Particle production correction}
\label{sec:particle}

Perturbing the hydrodynamic medium by depositing energy and momentum from energetic partons causes corrections to the final particle production. To illustrate the phenomenological consequences we recall the Cooper-Frye formula for the final particle distributions on a freeze-out surface $\Sigma_\mu (x)$ \cite{Cooper:1974mv}:%
\footnote{%
    The azimuthal angle $\phi_p$ associated with the momentum of an emitted particle should be distinguished from its spatial counterpart $\phi\equiv\arctan(x/y)$ for the azimuthal position of a fluid cell in coordinate space.
}
\begin{equation}\label{eq:distr}
    \frac{d^3N_i}{p_Tdp_Td\phi_pdy}=\frac{1}{(2\pi)^3}\int d^3\Sigma_\mu(x) p^\mu f_i(x, p)\,.
\end{equation}
Here $p^\mu$ is the four-momentum of the particle, $d^3\Sigma_\mu(x)$ the outward-pointing freeze-out surface normal vector for a surface element of volume $|d^3\Sigma_\mu|$ at point $x$, and $f_i(x, p)$ is the one-particle distribution function of species $i$, which for an ideal fluid in local thermal equilibrium takes the form
\begin{equation}
\label{eq:distr_f}
    \!\!\!\!
    f_i(x, p)= g_i \left[\exp{\left(\frac{u^\mu(x) p_\mu-Q_i\mu_f(x)}{T_f(x)}\right)} +\Theta_i\right]^{-1}\!\!.
\end{equation}
Here $i$ denotes the (in our case hadronic) particle species, $g_i$ its spin degeneracy factor, $\Theta_i\eq\pm1$ accounts for the fermionic ($+$) or bosonic ($-$) quantum statistics of the hadrons, $Q_i$ specifies their baryon charge, and $u^\mu(x)$ is the fluid flow velocity, $T_f(x)$ the freeze-out temperature and $\mu_f(x)$ the freeze-out chemical potential of the fluid at point $x$. We note that in principle for a dissipative fluid $f_i$ has viscous corrections which have here been ignored since we use an ideal fluid background.

The Lorentz invariant triple-differential momentum spectrum \eqref{eq:distr} can be integrated over $\phi_p$ or $p_T$ to yield the double-differential distributions $d^2N_i/p_Tdp_Tdy$ and $d^2N_i/d\phi_pdy$. We shall use \isd{} \cite{McNelis:2019auj} to compute these continuous particle distributions.\footnote{%
    Validations of the \bes{}\,+\,\isd{} framework using a setup similar to the one here can be found in L.~Du's Ph.D. thesis \cite{Du:2021fyr}.
}
We then subtract the final particle distribution obtained without the contributions from the energetic partons to obtain the particle production correction $\Delta N_i$ induced by partonic energy-momentum deposition into the fluid:
\begin{eqnarray}
\label{eq:pert_spectra_1}
    &&\frac{d^3\Delta N_i}{p_Tdp_Td\phi_pdy}=
\nonumber\\
    &&\left.\frac{d^3 N_i}{p_Tdp_Td\phi_pdy}\right|_\mathrm{w/\,parton}-\left.\frac{d^3 N_i}{p_Tdp_Td\phi_pdy}\right|_\mathrm{w/o\,parton}\!.
\end{eqnarray}
When the distortion caused by the partons can be treated as a small perturbation, the following expression can be used as a good approximation:
\begin{eqnarray}
\label{eq:pert_spectra_2}
    \frac{d^3\Delta N_i}{p_Tdp_Td\phi_pdy}\approx\frac{1}{(2\pi)^3}\int d\Sigma_\mu p^\mu \Delta f_i(x, p)\,.
\end{eqnarray}
Here $\Delta f_i(x, p)\equiv f_i(x, p)-f^{(0)}_i(x, p)$, with $f^{(0)}_i(x, p)$ being the one particle distribution on the unperturbed freeze-out surface $\Sigma_\mu$. Eq.~(\ref{eq:pert_spectra_2}) assumes a fixed freeze-out surface, i.e. that the freeze-out surface is not significantly shifted by adding the energy-momentum lost by the parton to the fluid. Fig.~\ref{fig:jet_illustration}b shows that, when energetic partons deposit energy and momentum into the medium, some hotter or colder regions are created. This in turn changes the space-time profile of the freeze-out surface as well as the hydrodynamic fields on it. Nonetheless the approximation \eqref{eq:pert_spectra_2} is frequently employed, especially in semi-analytical studies based on the linear response approach (see, e.g., Refs.~\cite{Yan:2017gmv, Yan:2017rku, Casalderrey-Solana:2020rsj}). 

In such studies, Eq.~(\ref{eq:pert_spectra_2}) is usually evaluated on a surface defined by a fixed ``freeze-out time'' $\tau_f$. To gain qualitative insight into the effect of hard partons plowing through the fluid on the final particle distributions one expands $\Delta f_i(x, p)$ as follows:
\begin{equation}
\label{eq:pert_spectra_3}
    \Delta f_i = f^{(0)}_i(1-\Theta_i f^{(0)}_i)[- u^\mu p_\mu \Delta\beta - \beta p_\mu\Delta u^\mu +Q_i\Delta\alpha],
\end{equation}
with $\beta=1/T$ and $\alpha=\mu/T$. This expression reveals that in a baryon-charged medium the parton-induced distortion of $\alpha$ causes, through the term $Q_i\Delta\alpha$, an additional difference in the spectra of protons and anti-protons, that supplements the differences seen in the spectra of other types of hadrons that do not carry baryon charge which are caused by the parton's effect on the temperature and flow profiles.

\subsection{Jet-induced vorticity}
\label{sec:vortex}

The energetic partons can also generate vortical structures in the medium and thus contribute to the polarization of the spins of particles emitted from the fireball \cite{Betz:2007kg, Becattini:2020ngo}. Jet-induced spin polarization in $A$-$A$ and $p$-$A$ collisions has attracted recent attention as a possible probe of shear viscosity and exotic flow structures such as ``smoke rings'' \cite{Lisa:2021zkj, Serenone:2021zef}. For spin-1/2 fermions in local thermal equilibrium, the ensemble average of the spin vector at space-time point $x$, to leading order in the thermal vorticity, reads \cite{Becattini:2013fla}
\begin{equation}
    S^{\mu}(x,p)=-\frac{1}{8m}\left(1-f_{F}(x,p)\right)\epsilon^{\mu\nu\rho\sigma}p_{\nu}\omega_{\rho\sigma}(x),\label{eq:spin_thermal}
\end{equation}
where $f_{F}(x,p)$ is the Fermi-Dirac distribution function and $\omega_{\rho\sigma}$ is the thermal vorticity defined by
\begin{equation}
    \omega_{\mu\nu} = \frac{1}{2}(\partial_\nu\beta_\mu-\partial_\mu\beta_\nu)\,,\label{eq:relativistic_vorticity_o}
\end{equation}
with $\beta^\mu(x) = u^\mu(x)/T(x)$ being the inverse temperature four-vector and $u^\mu(x)$ the fluid velocity. In Sec.~\ref{sec:vorticity}, we briefly discuss the vorticity induced by the energetic partons within a baryon-charged medium.

\section{Results and discussion}
\label{sec:results}

In this section we present results for the baryon-charged medium response to a single (Sec.~\ref{sec:singlejet}) and a back-to-back pair of energetic partons (Sec.~\ref{sec:dijet}) and discuss the resulting modifications of the emitted hadron distributions. At the end of Sec.~\ref{sec:dijet} we include a discussion of the phenomenological consequences of the $\mu$-dependence of the jet quenching coefficient $\qhat(T,\mu)$. Finally we explore in Sec.~\ref{sec:vorticity} the vortical pattern induced by a single energetic parton.  

\subsection{Medium response to a single energetic parton}
\label{sec:singlejet}

\begin{figure}[b]
    \centering
    \hspace{-0.45cm}\includegraphics[width= 0.51\textwidth]{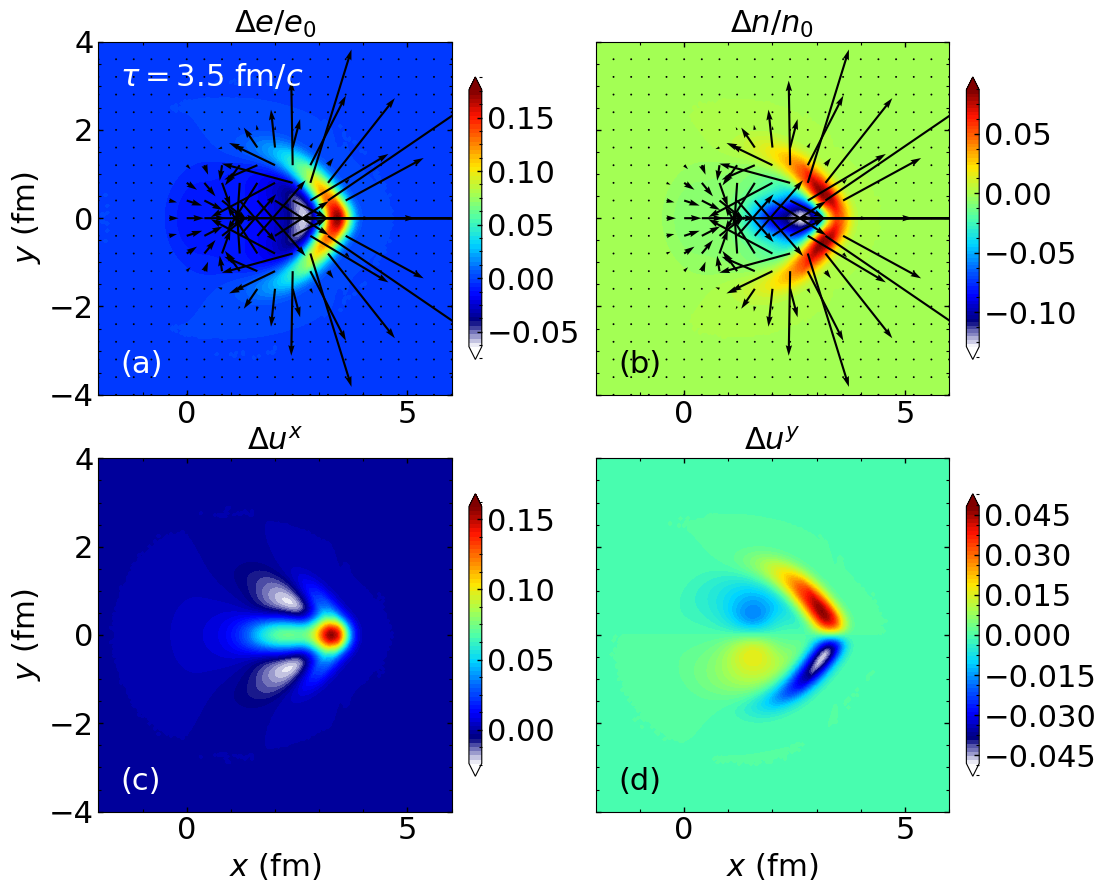}
    \caption{Transverse distributions of the perturbations in (a) the (rescaled) energy density $\Delta e/e_0$, (b) the (rescaled) baryon density $\Delta n/n_0$, and (c,d) the $x$- and $y$-components of the transverse flow, at $\tau{\,=\,}3.5$~fm$/c$, induced by the ``leading parton'' as shown by the right-pointing red arrow in Fig. \ref{fig:jet_illustration}a. The black arrows in panels (a,b) indicates the perturbation in the transverse flow vector, i.e. $\Delta \bm{u}_\perp = (\Delta u^x, \Delta u^y)$.}
    \label{fig:pert_transverse}
\end{figure}

In this subsection we study the effects of a single light-like parton starting at $x_0\eq1$\,fm (i.e. somewhat off-center, see Fig.~\ref{fig:jet_illustration}a) and moving in the $+x$ direction subject to energy loss described by the conformal $\qhat_\mathrm{CFT}$ given in Eq.~\eqref{eq:qcft}.\footnote{%
     The reader is invited to imagine this as the strongly interacting part of a $\gamma$-jet event in which the companion photon leaves the medium without interaction.}
Fig.~\ref{fig:pert_transverse} shows the distortions it induces in the medium for energy density (a), net baryon density (b), and for the two transverse flow components $u^x$ (c) and $u^y$ (d). For the energy and baryon number densities the jet-induced modifications are shown as fractions of the ideal fluid background (denoted by the subscript ``0") at each point.

In agreement with numerous earlier studies, one clearly recognizes a Mach cone in the perturbation of the energy density (Fig.~\ref{fig:pert_transverse}a). The hydrodynamic sound mode carries the energy deposited by the jet-parton into the conical region pointing to the right, dragging along some of the background fluid and leaving behind a depletion and trailing wake. The Mach cone opening angle $\phi_M\eq\sin^{-1}(c_s^2)$ (i.e., $\sin^{-1}(1/3)$ for our conformal EoS), is seen to be distorted by the radial flow profile of the background fluid \cite{Satarov:2005mv, Bouras:2014rea}. As can be seen qualitatively from the linearized equation \eqref{eq:hydro_source_pert_e}, the perturbation of the energy density, which causes a perturbation of the pressure gradients in the fluid, together with the momentum deposited by the jet-parton into the medium distort the transverse flow. This flow distortion is shown by the arrows in Figs.~\ref{fig:pert_transverse}a,b and as color contours plots in Figs.~\ref{fig:pert_transverse}c,d. Close inspection of the arrows in Figs.~\ref{fig:pert_transverse}a,b reveals a vortical pattern, reflecting the ``smoke rings'' discussed in Refs.~\cite{Betz:2007kg, Lisa:2021zkj, Serenone:2021zef}. We will discuss this in more detail in Sec.~\ref{sec:vorticity} below.

\begin{figure}[!tbp]
    \centering
    \hspace{-0.5cm}\includegraphics[width= 0.51\textwidth]{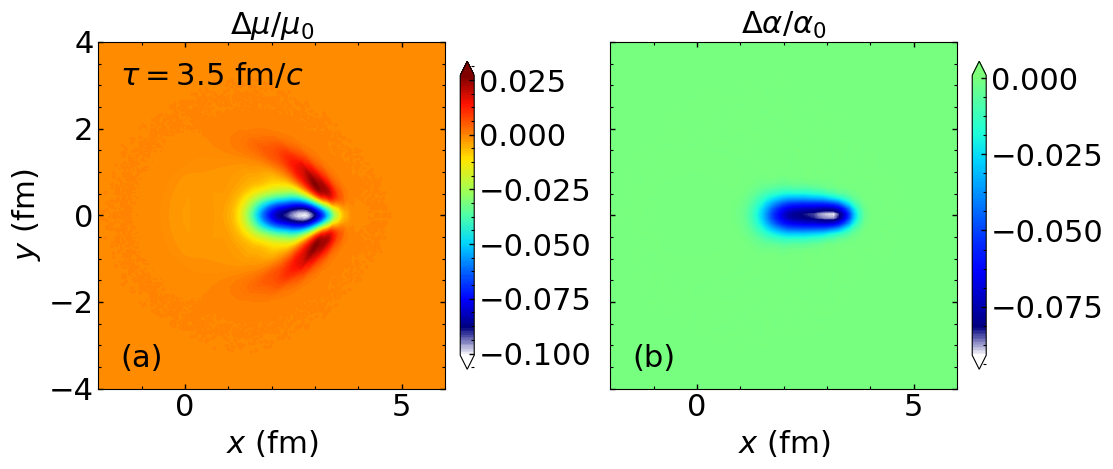}
    \caption{Same as Fig.\,\ref{fig:pert_transverse} but represented as fractional perturbations of (a) the baryon chemical potential $\mu$ and (b) the ratio $\alpha=\mu/T$ between the chemical potential and temperature.}
    \label{fig:pert_mu}
\end{figure}

A new feature arising in a baryon-charged fluid is a distortion of the background {\it baryon} density caused by the {\it energy-momentum} deposited by the jet. As we see in Fig.~\ref{fig:pert_transverse}b this phenomenon arises even in the absence of a baryon number transfer from the jet parton to the fluid. It is simply caused by the flow distortion resulting from the energy-momentum deposited into the background fluid since, in an ideal fluid, the net baryon number flows together with the momentum, without dissipation. As shown in Fig.~\ref{fig:pert_transverse}b, this results in a similar Mach cone structure for the net baryon density as we saw in Fig.~\ref{fig:pert_transverse}a for the energy density. A crucial difference is, however, the absence of a peak of the net baryon density at the location of the parton --- the Mach cone for the net baryon density has its wings but no apex. This is a consequence of baryon number conservation, since in our setup the parton does not add any baryon number to the medium. Baryon number can only be moved around in the fluid, and regions of increased net baryon density in Fig.~\ref{fig:pert_transverse}b (red) must be perfectly balanced by regions of decreased net baryon density (blue). (In panel (a) the positive and negative regions of $\Delta e$ do not balance because the jet provides net energy to the fluid.)\footnote{%
    This difference is also reflected in the asymmetric range of the color bar in panel (a) vs. the symmetric range in panel (b).}

Additional insight is provided by Fig.~\ref{fig:pert_mu} where we plot the net-baryon distortion in terms of distortions of the baryon chemical potential $\mu$ (panel (a)) and of its ratio $\alpha=\mu/T$ with the temperature (panel (b)). While the Mach cone structure in the net baryon distortion is reflected in the chemical potential, it is not visible in its ratio with $T$ whose distortion is basically zero, except for a narrow region along the parton's trajectory. This implies that in the cone region, where $\Delta\alpha=0$, $\mu$ and $T$ are distorted in the same way such that their ratio remains constant. Indeed, the fractional perturbation in the temperature, $\Delta T/T_0$, is very similar to that of the energy density (not shown). The similarity of the wings of the Mach cones for the energy and baryon density distortions illustrates the fact that both flow together, propagated via the sound mode, while along the parton's trajectory energy density gets sourced by the energetic parton whereas baryon density does not.

\begin{figure}[!tb]
    \centering
    \hspace{-0.45cm}\includegraphics[width=1.05\linewidth]{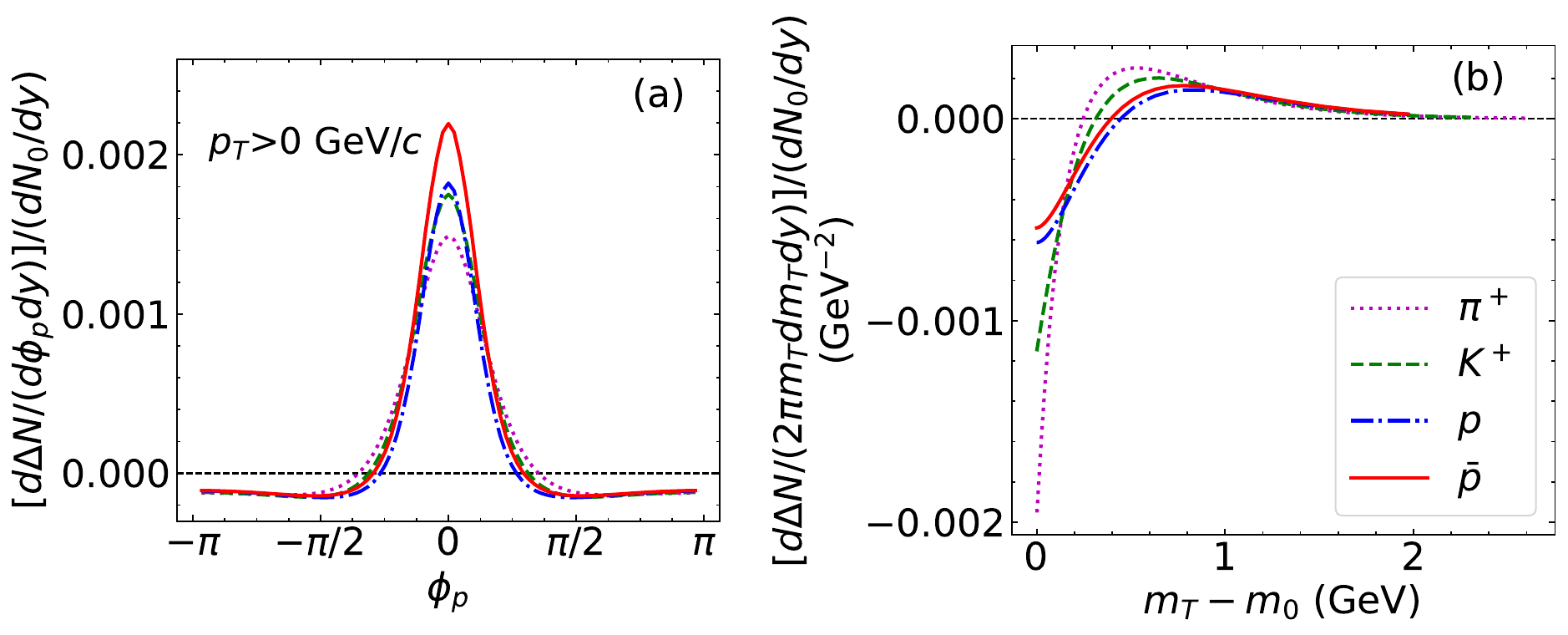}
    \caption{Particle production correction induced by a single energetic parton moving in the $+x$ direction, as a function of azimuthal angle $\phi_p$ (a) and of the transverse kinetic energy $m_T{-}m_0$ (b). For each particle species the distribution is rescaled by its background yield $dN_0/dy$ at mid-rapidity.
    \label{fig:pert_identified_phi_mT}
    }
\end{figure}

We now proceed to the phenomenological consequences of this hydrodynamic transport of the energy-momentum deposited by the jet parton in the medium, by evaluating its effects on the momentum spectra of the finally emitted particles. By integrating the distributions \eqref{eq:pert_spectra_1} over $p_T$ or $\phi_p$ we obtain the jet-induced modifications of the azimuthal and transverse mass ($m_T=\sqrt{m_0^2+p_T^2}$) distributions for pions, kaons, protons and anti-protons shown in Fig.~\ref{fig:pert_identified_phi_mT}. Note that, due to the azimuthal symmetry of Gubser flow, the $\phi_p$-dependent modifications shown in panel (a) sit on a flat background. For all hadron species we observe an enhancement of their yields emitted along the direction of the energetic parton --- a direct consequence of the deposited energy and momentum which increases the entropy propagating along $\phi\eq0$ and thus the number of particles emitted along $\phi_p\eq0$, broadened by thermal smearing which is inversely proportional to the hadron mass \cite{Schnedermann:1993ws}. We note that heavier particles experience a larger relative enhancement, and that a difference exists between protons and anti-protons. The latter reflects the non-zero chemical potential of the background medium: according to Eq.~\eqref{eq:pert_spectra_3}, $\Delta f_i$ receives a contribution ${\propto\,}Q_i\Delta\alpha$ which is positive for anti-protons and negative for protons because (as seen in Fig.~\ref{fig:pert_mu}b) $\Delta\alpha$ is negative. The enhancement at small $\phi_p$ is accompanied by a weaker depletion at larger $\phi_p$ which appears to be roughly the same for all hadron species and is spread out almost uniformly over the entire backward hemisphere opposite to the direction of the jet parton where one also finds the rarefaction wake.

Fig.~\ref{fig:pert_identified_phi_mT}b shows the corrections to the emitted particle distributions as a function of the transverse kinetic energy $m_T{-}m_0$. For all hadron species we observe a depletion at small transverse kinetic energy which turns into an enhancement for $m_T{-}m_0\gtrsim0.5\,$GeV. While the depletion at small $m_T{-}m_0$ and the location of the sign change exhibit strong mass dependence, the enhancement at large $m_T{-}m_0$ appears to be an almost universal function of transverse kinetic energy. This is qualitatively consistent with the correction being caused by hydrodynamic flow of the deposited energy-momentum which is known \cite{Schnedermann:1993ws} to break $m_T$-scaling at small transverse momentum. In addition, a small difference is observed between protons and anti-protons, reflecting the non-zero chemical potential of the background medium. 

\begin{figure}[!tbp]
    \centering
    \hspace{-0.5cm}\includegraphics[width=0.51\textwidth]{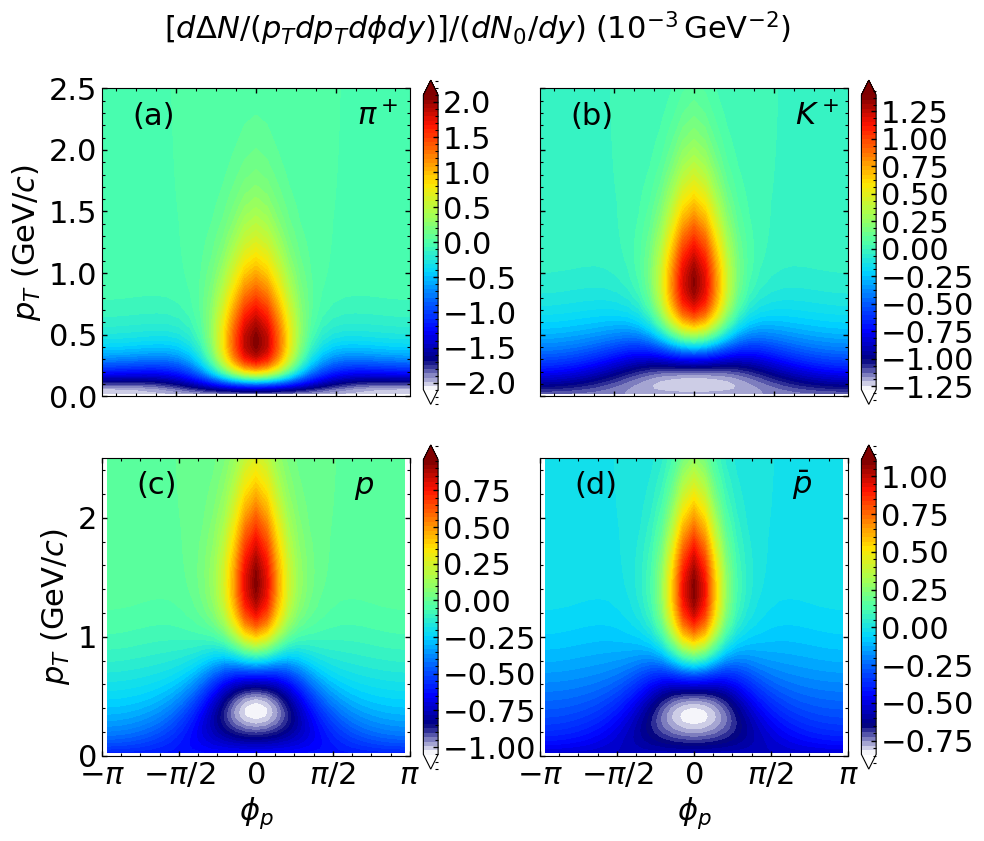}
    \caption{Two-dimensional distributions of the particle production correction in the $\phi_p{-}p_T$ plane, for identified pions ($\pi^+$, (a)), kaons ($K^+$, (b)), protons ($p$, (c)), and anti-protons ($\bar p$, (d)). The distributions are rescaled by the same background yields $dN_0/dy$ as in Fig.~\ref{fig:pert_identified_phi_mT}.
    \label{fig:pert_identified_phi_pT}
    }
\end{figure}

To better understand the results in Fig.~\ref{fig:pert_identified_phi_mT} we re-plot them in Fig.~\ref{fig:pert_identified_phi_pT} differentially as color contours in the $\phi_p$-$p_T$ plane. Near the jet axis, at $|\phi_p|\lesssim\pi/4$, {\it all} species are seen to be suppressed at low $p_T$ and enhanced at higher $p_T$. This suggests hydrodynamic flow of the deposited energy-momentum pushing particles from smaller to larger momenta along the direction of the jet parton, with heavier particles experiencing stronger flow boosts \cite{Schnedermann:1993ws}. In the $p_T$-integrated spectra of Fig.~\ref{fig:pert_identified_phi_mT} the flow-induced depletion at low $p_T$ is hidden. At large $\phi_p>\pi/2$, i.e. on the away-side of the jet parton, no enhancement is seen for any hadron species; as a result of the rarefaction wake, all hadron spectra are depleted. 

It is worth pointing out that in Fig.~\ref{fig:pert_identified_phi_mT} the jet-induced corrections to the hadron emission spectra are quite small. Still, rescaling the corrections shown in Fig.~\ref{fig:pert_identified_phi_mT}(a) by the unperturbed azimuthal distributions $dN_0/d\phi dy$ instead of $dN_0/dy$ increases these ratios by a factor $2\pi$ such that some of them reach a level of order 1\%. A similar-size jet-induced enhancement was previously observed in the CoLBT-hydro model, a hybrid approach coupling linear Boltzmann transport with hydrodynamics \cite{Chen:2017zte, Chen:2020tbl, Yang:2021qtl, Yang:2022nei}. That prediction qualitatively reproduces CMS data for both the azimuthal distributions and the soft hadron enhancement in the fragmentation functions \cite{CMS:2021otx}. More work is required to 
unambiguously associate these experimental findings with jet-induced Mach cones.
\subsection{Spectra distortions caused by an energetic dijet}
\label{sec:dijet}

We now proceed to considering a ``dijet'' event as depicted in Fig.~\ref{fig:jet_illustration}a in which, in addition to the ``leading jet parton'' moving in the $+x$ ($\phi\eq0$) direction where it will see a shorter medium and therefore lose less energy, there is a ``subleading jet parton'' emitted into the $-x$ ($\phi\eq\pi$) direction where it must traverse more matter and will lose more energy before escaping from the fireball medium. Given the smallness of the medium distortion effects caused by a single parton that we found in the preceding subsection, we simulate the medium response to the two partons separately and 
afterwards add their contributions to obtain the total response.\footnote{%
    We have studied the validity of this ``perturbative'' treatment by checking that, within the current setup, it produces final particle distributions that agree very accurately with those obtained from simulating the medium evolution with the source terms for both partons included simultaneously.\label{fn:pert}
}
We present only the corrections to the final particle distributions for pions and net protons; the former is a good qualitative substitute for the total charged hadron spectrum while the latter illustrates the specific effects caused by a non-zero net baryon charge in the background medium.

\begin{figure}[!tbp]
    \centering
    \hspace{-0.35cm}\includegraphics[width=\linewidth]{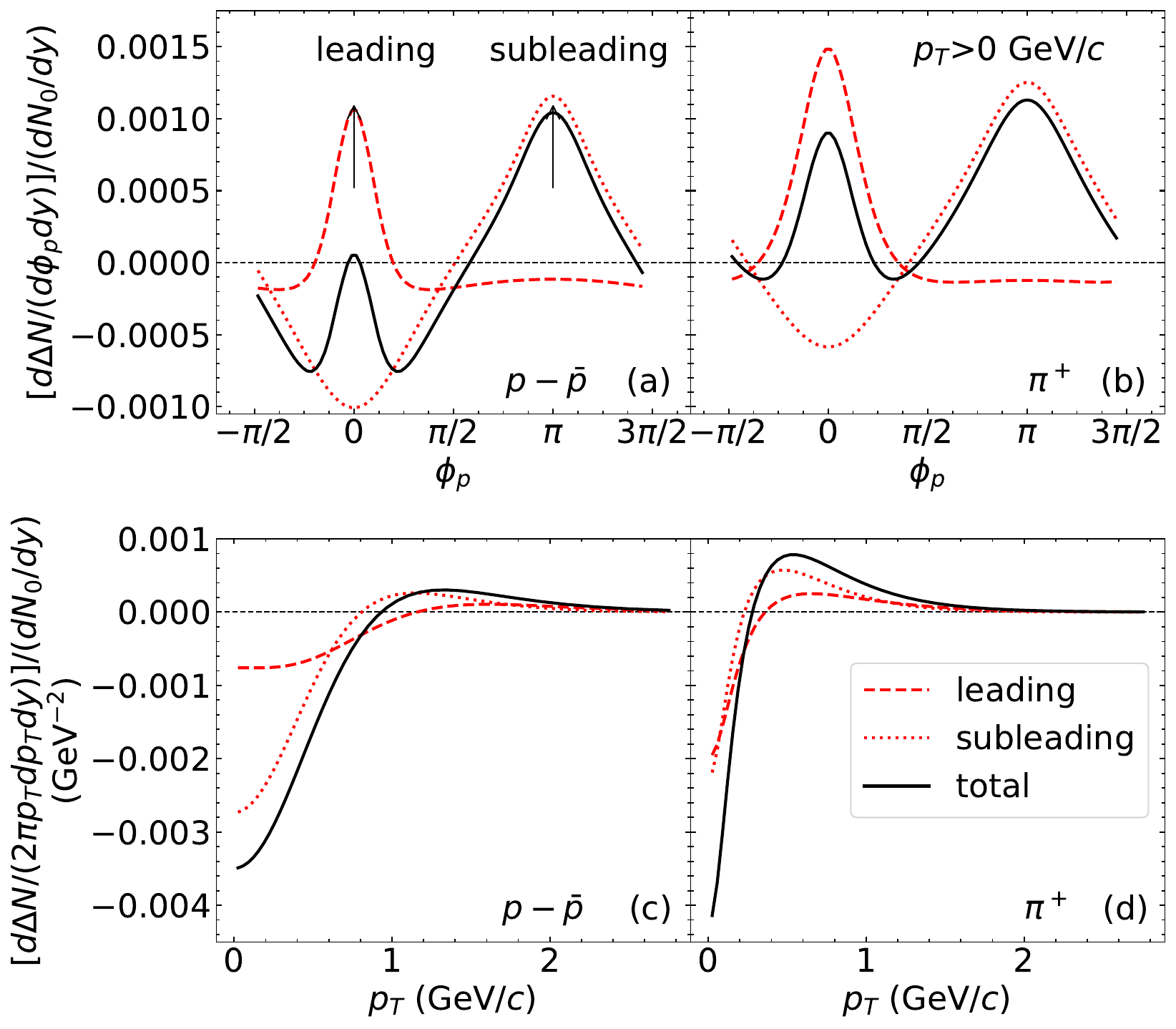}
    \caption{Particle production corrections induced by the energetic leading parton (red dashed line), the subleading parton (red dotted line), and their combined effect (black solid line). Azimuthal (a,b) and transverse momentum distributions (c,d) are shown for net protons (a,c) and pions.
    \label{fig:pert_dist_dijet}
    }
\end{figure}

The upper two panels in Fig.~\ref{fig:pert_dist_dijet} show the azimuthal distribution of the excess pions ($\pi^+$, panel (b)) and excess net protons ($p{-}\bar{p}$, panel (a)) induced by the energy loss of the leading parton (dashed red line, same as in Fig.~\ref{fig:pert_identified_phi_mT}), of the subleading parton (dotted red line), and of both of them together (solid black line). Since the parton pair is created off-center, at $x_0\eq1$\,fm, the subleading parton must first plow through the hot and dense fireball center before escaping on the other side (``away side'', $\pi/2{\,<\,}\phi{\,<\,}3\pi/2$). As the energy loss  $dE/dt\eq\ehat_\mathrm{CFT}\propto{}T^2$ grows quadratically with the temperature, the subleading parton deposits considerably more energy-momentum into the medium than the leading parton. This is clearly seen in Fig.~\ref{fig:pert_dist_dijet}b where the pion excess induced by the subleading parton on the away-side is significantly larger than the one induced by the leading parton on the ``near side'' ($-\pi/2{\,<\,}\phi{\,<\,} \pi/2$) (see also Refs.~\cite{Yan:2017rku,Yan:2017gmv}). Radial flow of the background medium collimates that excess; this explains the narrower width of the excess on the near side (where the deposited energy-momentum gets boosted forward by the background flow) than on the away side where the subleading parton first moves against the flow, then deposits a lot of energy near the fireball center where the radial flow is zero, and only after that feels a similar flow-induced forward boost as the leading parton. It is also interesting to observe that the large amount of energy deposited by the subleading parton near the center of the fireball leads to a stronger rarefaction wake behind its associated Mach cone, depleting the energy density behind it and causing a much larger reduction of the number of pions emitted on the near side near $\phi\eq0$ in response to the subleading parton (moving along $\phi\eq\pi$) than the reduction observed on the away side caused by the wake of the leading parton (which moves along $\phi\eq0$). When combined, the overall effect of the dijet is to create a double-humped pion excess (solid black line in Fig.~\ref{fig:pert_dist_dijet}b), with a smaller and narrower peak on the near side (i.e. in the direction of the leading parton) and a larger and wider peak on the away side (i.e. in the direction of the subleading parton).

\begin{figure}[!tbp]
    \centering
    \hspace{-0.35cm}\includegraphics[width=\linewidth]{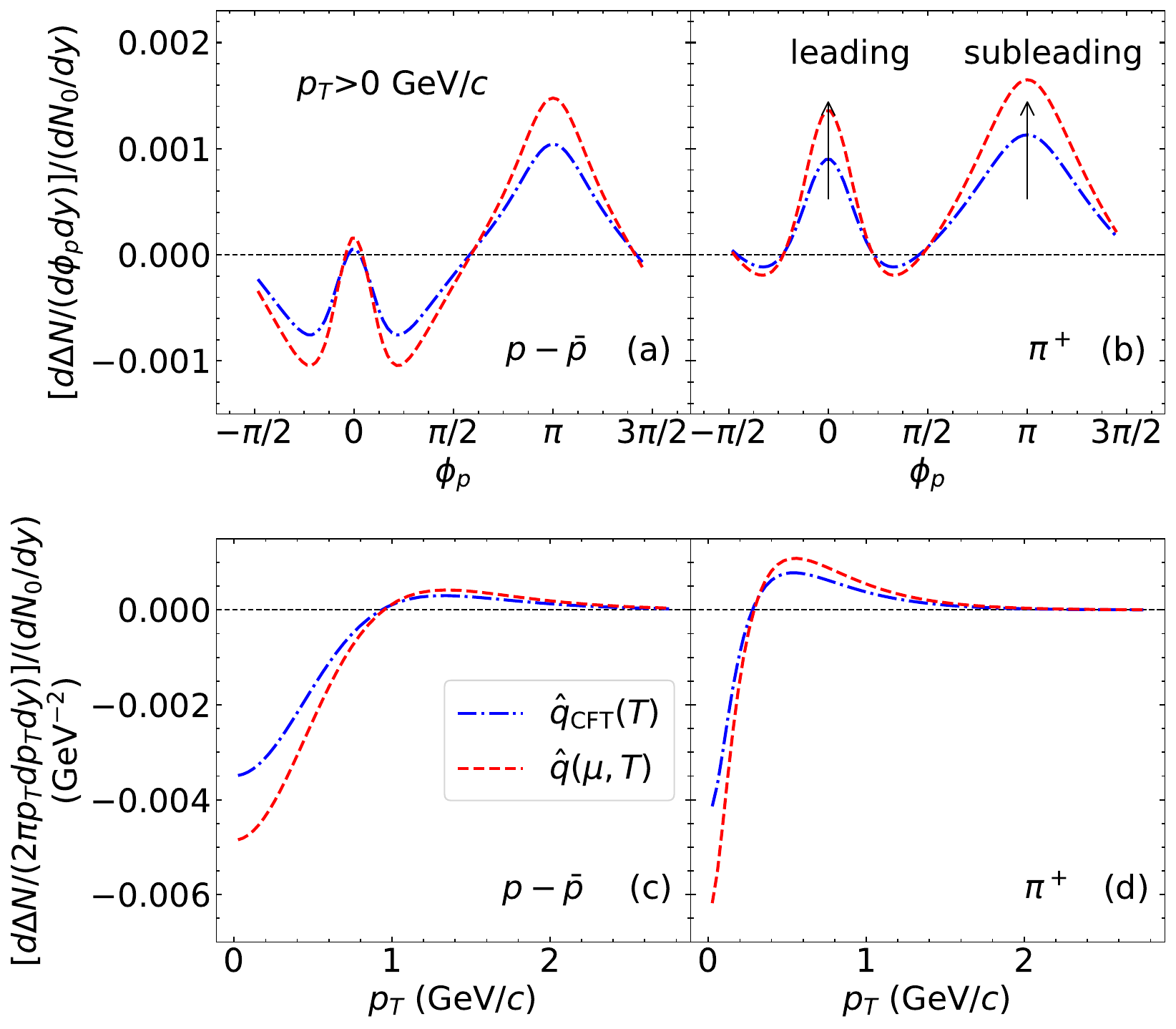}
    \caption{Same as Fig.~\ref{fig:pert_dist_dijet}, but showing only the combined effects from both partons in the dijet. The dash-dotted blue lines are identical with the solid black lines in Fig.~\ref{fig:pert_dist_dijet} and show the results for a conformal $\qhat_\mathrm{CFT}(T)$ that depends only on temperature. The red dashed lines show results for a non-conformal $\qhat(\mu, T)$.
    \label{fig:pert_dijet_cft}
    }
\end{figure}

The ``excess'' of net protons plotted in Fig.~\ref{fig:pert_dist_dijet}a shares many features with the pion excess in panel (b), except for the fact that the dijet deposits only net energy-momentum, but no net baryon number. Baryon number conservation manifests itself by a much stronger depletion of $p{-}\bar{p}$ on the near side (caused by the rarefaction wake generated by the subleading parton) than what is seen for pions in panel (b). When combining the (basically uncollimated) depletion of $p{-}\bar{p}$ from the wake of the subleading parton with the collimated enhancement of $p{-}\bar{p}$ caused by the leading parton, one obtains a double-humped net baryon depletion on the near side, as illustrated by the solid black line in Fig.~\ref{fig:pert_dist_dijet}a.

In the two bottom panels of Fig.~\ref{fig:pert_dist_dijet} we show the parton-induced corrections to the transverse momentum spectra of pions and net protons. We note that the low-$p_T$ suppression of net protons induced by the subleading parton is much stronger
than the one induced by the leading parton. The same does not happen for pions. This is consistent with our interpretation of that depletion being caused by the jet-induced modification of the hydrodynamic flow, which moves particles away from $p_T\eq0$ to larger $p_T$ (see Fig.~\ref{fig:pert_identified_phi_pT} above), and flow effects affecting heavier particles more strongly than lighter ones.

We close this subsection by briefly discussing the possible effects caused by an additional $\mu$-dependence of the jet quenching transport coefficient, $\qhat(\mu,T)$, such as the example shown in Fig.~\ref{fig:qhat}a. The discussion of Fig.~\ref{fig:qhat} revealed a generic tendency of a baryon chemical potential $\mu$ to increase the jet quenching parameter $\qhat(\mu,T)$, especially near the quark-hadron phase transition, giving rise to an increased energy loss rate. In Fig.~\ref{fig:pert_dijet_cft} we study the resulting effects on the modifications of the pion and net baryon $\phi_p$- and $p_T$-distributions. As expected, all the jet-induced medium effects discussed above are enhanced by moderate amounts (typically of order 50\%) when accounting for the $\mu$-dependence of $\qhat(\mu,T)$. Still, the jet-induced modifications of the emitted particle spectra remain very small. Using these observables to provide compelling experimental support for a critical increase of $\qhat(\mu,T)$ near the quark-hadron phase transition \cite{Liao:2008dk} will not be easy.

\subsection{Parton induced vorticity}
\label{sec:vorticity}

The unperturbed medium undergoing Gubser expansion is boost-invariant and has azimuthal symmetry and thus no vorticity. But, as illustrated in Fig.~\ref{fig:pert_transverse}, by depositing energy and momentum into the medium, an energetic parton traversing that medium induces non-trivial perturbations in the temperature profile and the transverse flow. Since our set-up preserves longitudinal boost-invariance, the vorticity around any axis perpendicular to the beam direction $z$ remains zero, but jet-induced modifications of the transverse flow and transverse temperature gradients can induce an interesting vortical pattern around the beam axis. It is the remnant of a vortex ring (``smoke ring'' \cite{Betz:2007kg, Lisa:2021zkj, Serenone:2021zef}) generated by a jet at $z\eq0$ escaping the fireball in $x$ direction, after cutting that ring by the $z\eq0$ plane and extending the resulting pattern boost-invariantly along the $z$ direction.

With the background medium having zero vorticity, the parton induced vorticity can be written in terms of the parton-induced changes in the flow pattern, $\Delta u_\mu$, and in the temperature gradients, $\partial_\mu\Delta T$. To separate these contributions
we rewrite the (unitless) thermal vorticity tensor  \eqref{eq:relativistic_vorticity_o} and split it into two terms \cite{Becattini:2015ska}:
\begin{eqnarray}
\label{eq:vort-split}
    \omega_{\mu\nu}\equiv\frac{1}{T}\omega_{\mu\nu}^\mathrm{(k)}+\omega_{\mu\nu}^\mathrm{(th)}(T)\,.
\end{eqnarray}
Here 
\begin{eqnarray}
\label{eq:omega_k}
    \omega_{\mu\nu}^\mathrm{(k)}=-\frac{1}{2}\left(\partial_\mu \Delta u_\nu - \partial_\nu \Delta u_\mu\right)\,,
\end{eqnarray}
is the so-called {\it kinetic vorticity} associated with vortical flow structures and 
\begin{eqnarray}
\label{eq:omega_th}
    \omega_{\mu\nu}^\mathrm{(th)}(T)=-\frac{1}{2T^2}\left(u_\mu\partial_\nu{}\Delta T - u_\nu\partial_\mu{}\Delta T\right)
\end{eqnarray}
captures the contribution from temperature gradients.

\begin{figure}[!tbp]
    \centering
    \hspace{-0.5cm}\includegraphics[width=0.51\textwidth]{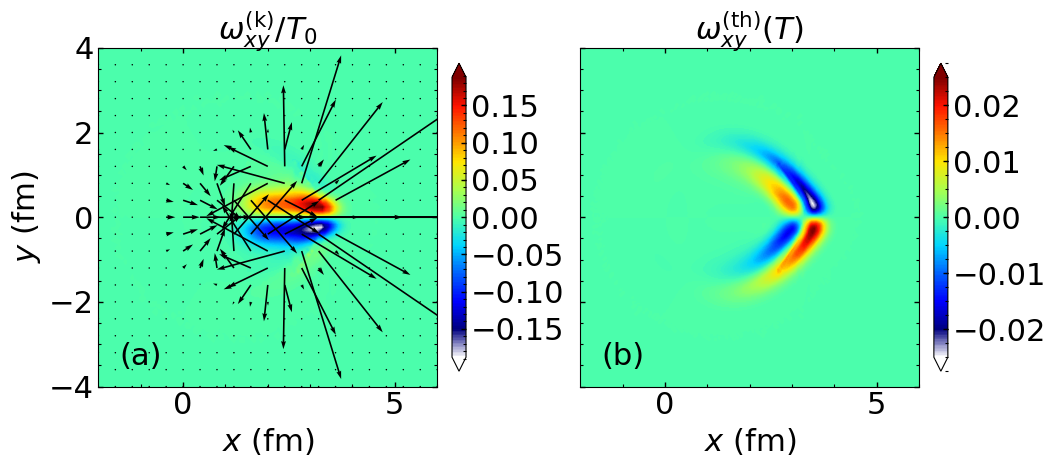}
    \caption{Transverse distribution of (a) the kinetic vorticity (\ref{eq:omega_k}) and (b) its temperature gradient contribution (\ref{eq:omega_th}), at $\tau{\,=\,}3.5$~fm$/c$. The arrows in panel (a) are the same as in Fig.~\ref{fig:pert_transverse}, indicating the perturbation in the transverse flow which is the root cause of the jet-induced vorticity.
    \label{fig:dist_vorticity}
    }
\end{figure}

The $xy$ components of these two vorticity tensors (corresponding to the $z$ component of the associated vorticity vector), evaluated with the temperature and flow profiles associated with the distributions shown for $\tau\eq3.5$\,fm$/c$ in Fig.~\ref{fig:pert_transverse}, are shown in Fig.~\ref{fig:dist_vorticity}. In panel (a) we illustrate the transverse flow modification $\Delta \bm{u}_\perp=(\Delta u^x, \Delta u^y)$ as before by black arrows. A vortical structure is clearly visible, corresponding to a non-vanishing kinetic vorticity $\omega_{xy}^\mathrm{(k)}$. The magnitude and sign of the latter is represented by color in the same plot. The kinetic vorticity is seen to be antisymmetric (same magnitude, opposite sign) under reflection $y\to-y$, i.e. left and right of the parton's trajectory. 

The contribution from thermal gradients, $\omega_{\mu\nu}^\mathrm{(th)}(T)$, is shown in panel (b). It is non-zero mainly along the two wings of the Mach cone where the energy density and thus the temperature is enhanced in response to the energy-momentum deposited by the energetic parton (see Fig.~\ref{fig:pert_transverse}a). Since $\omega_{\mu\nu}^\mathrm{(th)}(T)$ involves the gradient of $\Delta T$, each wing from $\Delta T$ contributes a pair of wings with opposite signs to the color plot of $\omega_{\mu\nu}^\mathrm{(th)}(T)$. Like its kinetic counterpart, it is antisymmetric under $y\to -y$. At the front edge of the Mach cone it contributes with the opposite sign of the kinetic vorticity, at its rear edge it adds to the kinetic vorticity with the same sign. In magnitude the first term in Eq.~(\ref{eq:vort-split}) (i.e. the contribution from the flow modification) dominates over the second term (arising from the temperature modification) by about an order of magnitude; in other words, the parton-induced vorticity distributes mostly in the direct vicinity of the parton's path and not so much over the wings of the Mach cone.

We note that the magnitude of the parton-induced thermal vorticity plotted in Fig.~\ref{fig:dist_vorticity} roughly follows the magnitude profile of $\Delta\alpha$ shown in Fig.~\ref{fig:pert_mu}b. As discussed earlier, $\Delta\alpha$ is the root cause for the observed differences in particle emission between baryons and anti-baryons. Combined with the induced thermal vorticity this can induce differences in the momentum distribution of the polarization between $\Lambda$ and $\bar\Lambda$ hyperons emitted from the fireball. Methods for measuring the smoke-ring-shaped jet-induced hyperon polarization were recently proposed in Refs.~\cite{Lisa:2021zkj, Serenone:2021zef}.

\section{Conclusions}
\label{sec:conclusion}

We explored qualitative features of the baryon-charged medium response to energetic partons, using the \bes{} + \isd{} code package. In order to focus on the essential features of the jet-induced medium response, we studied their properties on a simplified background medium undergoing ideal Gubser flow with a conformal (massless) equation of state. Energy-momentum deposition by the energetic partons into the medium is controlled by the energy loss rate $\ehat$ and described by dynamical source terms in the hydrodynamic equations. Using the fluctuation-dissipation theorem to relate $\ehat$ to the ``jet-quenching" (more accurately: transverse broadening) parameter $\qhat$, we employed a functional dependence $\qhat(\mu, T)$ where, in a baryon-charged medium, the latter depends on both temperature and baryon chemical potential. While our approach ignores the additional complications of event-by-event fluctuations in the initial temperature and chemical potential profiles, as well as the effects of jet fragmentation (i.e. the splitting of a single energetic parton into multiple particles), it offers conceptual clarity and focus.

We ignored the discrete exchange of baryon number between the jet parton and the medium, focusing on the continuous exchange of energy and momentum. For single partons as well as ``di-jets'' consisting of two back-to-back partons, we explored the hydrodynamical propagation of the deposited energy-momentum and the resulting distortions in energy density, net baryon density, and the hydrodynamic flow profile. We saw that the medium response is associated with Mach-cone-like structures in the energy and baryon density, temperature and chemical potential, but not the ratio $\alpha=\mu/T$. This last feature is explained by the lack of baryon-doping from the jet parton, combined with the conformality of our EoS and our assumed lack of dissipation in the background medium, which (via the law of baryon number conservation) implies identical hydrodynamic flows for energy-momentum and net baryon number. In a conformal ideal fluid that evolves isentropically $\mu/T$ is constant along the expansion trajectories, and as the heat caused by the deposited energy propagates outward in a Mach cone shaped sound wave, the baryon chemical potential follows along. Only in the immediate vicinity of the energetic parton's trajectory, where energy is added to the fluid but no baryon number, is the ratio $\mu/T$ visibly affected.  
    
Using the Cooper-Frye algorithm we computed the modifications caused by the medium response of the identified hadron spectra emitted from an isothermal freeze-out surface. We identified peak-like structures in the azimuthal angular distribution caused by the energetic parton(s) and discussed their specific features in the context of the Mach-cone-shaped compression waves and the trailing rarefaction wake induced in the medium. We also observed that the hydrodynamic flow induced by the jet parton(s) moves hadrons to larger transverse momenta, more strongly so for the heavier baryons than the lighter mesons, leading to a depletion at small $p_T$ and an enhancement at larger $p_T$. We found that net-baryon conservation in the fluid medium leads to different manifestations of these phenomena for baryon-charged (net) protons and for uncharged mesons. We also noted that the dependence of $\qhat$ on the baryon chemical potential causes an increase of the parton energy loss in baryon-charged fluids, especially in the vicinity of the quark-hadron phase transition, increasing all observed medium response effects by typically 50\%. Nevertheless, none of the medium effects and spectra modifications caused by hydrodynamic response to the energetic parton(s) exceeded about a percent of the background fluid properties. While our study confirms the in-principle observability of identified particle signatures of jet-induced Mach cones, measuring them experimentally with sufficient precision for an unambiguous theoretical confirmation of their origin continues to be a challenge.

Finally, we studied the thermal vorticity generated in the fluid by the medium response. We found that induced flow effects dominate by about an order of magnitude over contributions to the thermal vorticity arising from additional temperature gradients in response to the deposited energy-momentum. The structures observed in this work could be interpreted as the vestiges of the ``smoke rings" recently observed in Refs.~\cite{Lisa:2021zkj, Serenone:2021zef}. These rings surround the trajectory of the energetic parton as it plows through the medium.

%
\section*{Acknowledgements}
%
Insightful comments by Xin-Nian Wang on the first draft of this work after it was posted online are gratefully acknowledged. We also thank Chandrodoy Chattopadhyay, Derek Everett, Mike McNelis, Chanwook Park, Shuzhe Shi, Yasuki Tachibana and Xiaojun Yao for fruitful discussions. L.D. expresses his gratitude to Rômulo Rougemont for providing tabulated results for $\qhat$ from Ref.~\cite{Rougemont:2015wca}. This work was supported in part by the U.S. Department of Energy (DOE), Office of Science, Office for Nuclear Physics under Award No.\ \rm{DE-SC0004286} and within the framework of the BEST Collaboration, by the National Science Foundation (NSF) within the framework of the JETSCAPE Collaboration under Award No.\ \rm{ACI-1550223}, and by the Natural Sciences and Engineering Research Council of Canada. U.H. acknowledges support from a Research Prize from the Alexander von Humboldt Foundation. Computing resources were generously provided by the Ohio Supercomputer Center \cite{OhioSupercomputerCenter1987} (Project PAS0254).
%

\bibliography{jetmedium}

\end{document}